\documentclass[12pt]{iopart}

\usepackage{graphicx}
\expandafter\let\csname equation*\endcsname=\relax 
\expandafter\let\csname endequation*\endcsname=\relax
\usepackage{amsmath,amssymb}
\usepackage{multirow}
\usepackage[usenames,dvipsnames]{xcolor}
\usepackage{mathtools}
\usepackage{gensymb}
\usepackage{physics}
\usepackage{siunitx}
\usepackage{hyperref}
\usepackage[style=phys]{biblatex}

\definecolor{ggreen}{rgb}{0.0, 0.5, 0.0}

\newcommand{\omu}{Department of Mechanical Engineering, Osaka Metropolitan University, 1-1 Gakuen-tyo, Naka-ku, Sakai, Osaka 599-8531, Japan}

\newcommand{\rev}[1]{\textcolor{black}{#1}}

\begin{document}

\title[Two apparently different interfacial stress formulations in the DI models]{Two \rev{apparently different} interfacial stress formulations constructed from the diffuse interface free energy model}

\author{Takeshi Omori}
\ead{t.omori@omu.ac.jp}
\address{\omu}

\begin{abstract}
To express the capillary stress in the diffuse interface method, there are two different formulations in the literature: one formulation is proportional to the density and the other is to the density gradient. Confusingly, these two apparently different formulations both have been widely used for the same purpose without much attention in the difference. In the present study, we theoretically show that only the latter represents the capillary stress and the former represents the stress due to the non-equilibrium irreversible process in the fluid. The formulations are analyzed not only for the one-component fluid but also for the multi-component fluid. In the prediction of the incompressible flow of one-component fluids, the two formulations turn out to give the identical velocity field, but not the pressure field even in this case.
\end{abstract}

%
%
%

\section{Introduction}
Fluid flows with interfaces between distinct phases\rev{/components} are encountered in many scientific and industrial occasions, and the prediction of these flows have been pursed for the latest decades \cite{Scardovelli1999,Anderson1998,anthony2023}. Modeling the interfacial tension and its appropriate form as a capillary stress term that is added to the Navier-Stokes (NS) equation, namely the force per unit fluid volume due to the surface tension, is one of the key issues to predict the interfacial flows. The formulation of the capillary stress term depends on the choice of the advection scheme of the marker function, which is a function to identify the different fluids in the system and track the motion of the interface. The diffuse-interface (DI) method, sometimes also called the phase-field method as called in the metallurgy community, is one of such advection schemes of the marker function. The DI method has its origin in the thermodynamics of solutions \cite{cahn1958}, which is a distinct feature separating the DI method from the other purely numerical schemes such as the volume-of-fluid, the level-set, \rev{the cubic-interpolated pseudoparticle (CIP)} and the front-tracking methods \cite{Tryggvason2011}. Because of its theoretical background, the DI method is considered to be particularly suitable to describe the flows where the characteristic length scale is comparable to the interfacial thickness, e.g. near critical interfacial phenomena, flows with moving contact lines, and breakup and coalescence of droplets \cite{Anderson1998}. 

The DI method stands on the free energy model of inhomogeneous systems. The system free energy $F$ is typically modeled by
\begin{equation}\label{eq:F}
F = \int \dd{\vb{r}}
\left[
f_0(\phi) + \frac{K}{2}(\grad{\phi})^2
\right]
\end{equation}
\rev{for one-component systems (The system free energy for multi-component systems is given in Sec.~\ref{sec:osmotic})}. The variable $\phi$ can be the mass/molar fraction of one solute component in a solution \cite{masaodoi2013,cahn1958}, or the statistically averaged number density of fluid molecules of one component \cite{Stephan2019} in the system, which is used as the marker function in the DI method. We employ here the latter definition for $\phi$ since \rev{it is more convenient} in capturing the interface between the different phases in the fluid. The homogeneous contribution to the free energy density away from the interface is denoted as $f_0$ in Eq.~\eqref{eq:F}. The second term on the RHS of Eq.~\eqref{eq:F} with the positive constant $K$ is essential to model the excess free energy of the interface and the interfacial tension $\sigma$ with respect to $\phi$ given \cite{Hansen2013} by 
\begin{equation}\label{eq:st}
\sigma = \int \dd{\xi} K \left(\dv{\phi}{\xi}\right)^2,
\end{equation}
where $\xi$ is the coordinate normal to the interface. 
The marker function $\phi$ is updated by the following equation called the Cahn-Hilliard (CH) equation for incompressible flows:
\begin{equation}\label{eq:ch}
\pdv{\phi}{t}+\vb*{u}\vdot\grad{\phi}=
M\laplacian(\fdv{F}{\phi}),
\end{equation}
where $\vb*{u}$ is the fluid velocity, $\fdv*{}{\phi}$ denotes the functional derivative \rev{with respect to $\phi$}, and $M$ determines how fast the system transitions to the equilibrium state having zero net driving force by $\nabla(\fdv*{F}{\phi})$. 

As the capillary stress term 
added to the NS equation
, two different DI formulations exist in the literature. One formulation is proposed \cite{Jacqmin1996} as 
\begin{equation}\label{eq:jacqmin}
-\phi\grad{\mu},
\end{equation}
and the other \cite{Chella1996,Qian2003} is
\begin{equation}\label{eq:chella}
\mu\grad{\phi},
\end{equation}
where $\mu\equiv\fdv*{F}{\phi}$. The quantity $\mu$ is different from the chemical potential\footnote{The quantity $\mu$ is often simply called the chemical potential \cite{Chella1996,Jacqmin1999,Qian2006a}, but it is not the true thermodynamic chemical potential, which is constant throughout the system in equilibrium \cite{widom1978}.} by $-K\laplacian{\phi}$ \cite{omori2024}. These two apparently different formulations both have been widely used \rev{as a capillary stress term} without much attention in the difference. 
In the present study, we show that the former of the two formulations \rev{is} generated from the free energy [Eq.~\eqref{eq:F}] by Onsager's variational principle \cite{Onsager1931a,Onsager1931}, representing the \rev{non-equilibrium} irreversible process in the fluid. This formulation [Eq.~\eqref{eq:jacqmin}] does not \rev{include} the interface curvature \rev{information}, meaning that it does not represent the capillary stress term \rev{that is directly linked to the Laplace pressure}. On the other hand, the latter formulation\rev{, which is derived from the equilibrium mechanical balance in the system,} does represent the capillary stress term. 

\section{Stress formulation in the fluid under irreversible processes}\label{sec:osmotic}
We \rev{start by considering that the fluid is composed of multiple components and the one-component case follows}. 
The temporal variation of the density of one component \rev{$\alpha$} of interest is described by the mass conservation law
\begin{equation}\label{eq:cont}
\pdv{\phi_{\rev{\alpha}}}{t}+\div(\vb*{v}_{\rev{\alpha}} \phi_{\rev{\alpha}})=0,
\end{equation}
where the \rev{component} velocity $\vb*{v}_{\rev{\alpha}}$ can be different from the fluid velocity $\vb*{u}$. We assume that the fluid flow is incompressible, $\div{\vb*{u}}=0$. If we write $\vb*{J}_{\rev{\alpha}}=(\vb*{v}_{\rev{\alpha}}-\vb*{u})\phi_{\rev{\alpha}}$, Eq.~\eqref{eq:cont} can be rewritten as
\begin{equation}\label{eq:phi}
\pdv{\phi_{\rev{\alpha}}}{t}+\div(\vb*{u}\phi_{\rev{\alpha}})=-\div{\vb*{J}_{\rev{\alpha}}}.
\end{equation}
\rev{%
The system free energy $F$ is modeled \cite{Davis1982,omori2024} as
\begin{equation}\label{eq:F_multi}
F = \int \dd{\vb{r}}
\left[
f_0(\vb*{\phi}) + 
\sum_{\alpha,\beta}\frac{K_{\alpha\beta}}{2}\grad{\phi_{\alpha}}\vdot\grad{\phi_{\beta}}
\right],
\end{equation}
where $\vb*{\phi}=[\phi_1,\phi_2,\cdots]^T$ is the vector of component densities. 
}
According to Onsager's variational principle \cite{Onsager1931a,Onsager1931}, which is a general framework to describe the irreversible processes \cite{doi2011}, the fluid system should behave minimizing the Rayleighian $R$ given by
\begin{equation}\label{eq:rayleighian}
R=\frac{1}{2}W+\pdv{F}{t}-\int\dd{\vb{r}}p\div{\vb*{u}}.
\end{equation}
The last term on the RHS is the constraint for the incompressibility and $p$ is the Lagrange multiplier. 
The energy dissipation $W$ has two origins: one is the viscous motion of the fluid and the other is the relative motion between the fluid component and the background fluid. The former is written as
\begin{equation}\label{eq:w1}
W_1=
\frac{1}{2}\int\dd{\vb{r}}
\eta
\left(
\grad{\vb*{u}}+\grad{\vb*{u}}^T
\right)^2
\end{equation}
and the latter we model \cite{Qian2006a} as
\begin{equation}\label{eq:w2}
W_2=
\int\dd{\vb{r}}\frac{\vb*{J}_{\rev{\alpha}}^2}{M_{\rev{\alpha}}},
\end{equation}
where $\eta$ is the fluid viscosity and $M_{\rev{\alpha}}$ is the reciprocal of the friction coefficient regarding $\vb*{J}_{\rev{\alpha}}$. Taking the variation of $R$ with respect to $\vb*{u}$ as zero, we obtain (See \ref{sec:onsager_detail} for the complete derivation)
\begin{equation}\label{eq:stress_onsager}
\div{
\left\{
\eta\left[
\grad{\vb*{u}}+(\grad{\vb*{u}})^T
\right]
\right\}
}
-\grad{p}
-\rev{\sum_{\alpha}}\phi_{\rev{\alpha}}\grad{\mu_{\rev{\alpha}}}
=0,
\end{equation}
and taking the variation of $R$ with respect to $\vb*{J}_{\rev{\alpha}}$ results in 
\begin{equation}\label{eq:J}
\vb*{J}_{\rev{\alpha}}=-M_{\rev{\alpha}}\grad{\mu_{\rev{\alpha}}}.
\end{equation}
From Eqs.~\eqref{eq:phi} and \eqref{eq:J} we recover the CH equation \rev{for the multi-component system} assuming $M_{\rev{\alpha}}$ is constant in space:
\begin{equation}\label{eq:phi_multi}
\rev{%
\pdv{\phi_{\alpha}}{t}+\vb*{u}\vdot\grad{\phi_{\alpha}}
=M_{\alpha}\laplacian{\mu_{\alpha}}.
}
\end{equation}

Equation \eqref{eq:stress_onsager} is the equation of motion of the fluid with inhomogeneous $\rev{\vb*{\phi}}$. The unsteady and advection terms $\mathrm{D}\vb*{u}/\mathrm{D}t$ can be understood as the imbalance of the stress terms in Eq.~\eqref{eq:stress_onsager} to form the full NS equation. The last term on the LHS is the interface related term, \rev{whose physical meaning should be sought here}. Since the osmotic pressure $\Pi_{\rev{\alpha}}$ \rev{concerning the component $\alpha$} can be related to the free energy density \cite{masaodoi2013} by\footnote{\rev{The independence between the osmotic pressures concerning different components is assumed.}} 
\begin{equation}
\grad{\Pi_{\rev{\alpha}}}=\phi_{\rev{\alpha}}\grad(\pdv{f_0}{\phi_{\rev{\alpha}}}),
\end{equation}
the last term on the LHS of Eq.~\eqref{eq:stress_onsager} \rev{can be written as}
\begin{equation}\label{eq:decomp}
\rev{%
-\sum_{\alpha}\phi_{\alpha}\grad{\mu_{\alpha}}
=
-\sum_{\alpha}\grad{\Pi_{\alpha}}+
\sum_{\alpha,\beta}\frac{K_{\alpha\beta}}{2}\phi_{\alpha}\grad\laplacian{\phi_{\beta}}.
}
\end{equation}
\rev{None of the terms on the RHS include the interface curvature information and hence do not represent the capillary stress.} It is reasonable that the capillary stress term, which does not represent non-equilibrium irreversible phenomena, is not generated from the free energy model by Onsager's variational principle. 

\rev{%
When the system consists of only one component, Eqs.~\eqref{eq:stress_onsager} and \eqref{eq:J} reduce to 
\begin{equation}\label{eq:stress_onsager_one}
\div{
\left\{
\eta\left[
\grad{\vb*{u}}+(\grad{\vb*{u}})^T
\right]
\right\}
}
-\grad{p}
-\phi\grad{\mu}
=0
\end{equation}
and
\begin{equation}\label{eq:J_one}
\vb*{J}=-M\grad{\mu},
\end{equation}
respectively. However, when there is only one component in the system, $\vb*{J}=0$ by construction. It requires $\grad{\mu}=0$ by Eq.~\eqref{eq:J_one}. For the one-component system, the CH equation \eqref{eq:ch} therefore reduces to the continuity equation for incompressible flows:
\begin{equation}
\pdv{\phi}{t}+\vb*{u}\vdot\grad{\phi}=0.
\end{equation}
}

\rev{It is worthwhile to compare the present results with the other formulations in the literature based on  thermodynamics. Jacqmin \cite{Jacqmin1996} obtained the same expression $-\phi\grad{\mu}$ considering the total energy in the system to be constant, where the total energy includes the kinetic energy of the fluid.} The Korteweg-type stress, \rev{which can be interpreted as the stress to preserve the homogeneity of the free energy density under the constraint of mass conservation}, has been suggested \cite{Anderson1998} as the interfacial stress:
\begin{equation}
\mathbb{T}=
\left[
f_0(\phi)+\frac{K}{2}(\grad{\phi})^2-\phi\mu
\right]\mathbb{I}
-
K\grad{\phi}\otimes\grad{\phi}.
\end{equation}
\rev{In the NS equation, the divergence of this stress ($\div{\mathbb{T}}$) should be evaluated.} Since $\div{\mathbb{T}}=-\phi\grad{\mu}$, \rev{the function of the Korteweg-type stress in the NS equation} is in fact equivalent to the last term on the LHS of Eq.~\eqref{eq:stress_onsager_one}. \rev{In contrast to the present study, the mechanism to show $\grad{\mu}=0$ has been missing due to the ignorance of the irreversible process by the fluid flow.}

\section{\rev{Stress formulation in the fluid in equilibrium}}
\rev{It was shown in the previous section that the capillary stress is not involved in the non-equlibrium irreversible processes, and then it should be derived from the equilibrium mechanical balance between two parts of the fluid separated by the interface. From the mechanical balance, the capillary stress term in the NS equation can be written as \cite{Tryggvason2011}
\begin{equation}\label{eq:capillary}
\sigma\kappa\vb*{\xi}\delta(\xi),
\end{equation}
where $\kappa$ is the interface curvature, $\vb*{\xi}$ the unit normal vector to the interface, and $\delta$ the Dirac delta function. Equation \eqref{eq:capillary} can be related to the free energy and the density variation in the system \cite{Chella1996,liu2014}.
}

\rev{We consider first the one-component case for simplicity. }Suppose the interfacial unit normal $\vb*{\xi}$ pointing to the direction of increasing $\phi$, $\mu$ \rev{($\equiv\fdv*{F}{\phi}$)} can be rewritten as
\begin{equation}\label{eq:mu}
\mu=
\dv{f_0}{\phi}
-K\pdv[2]{\phi}{\xi}-K\pdv{\phi}{\xi}\div\vb*{\xi}
\end{equation}
since $\grad{\phi}=(\pdv*{\phi}{\xi})\vb*{\xi}$ and $\laplacian\phi=\pdv*[2]{\phi}{\xi}+(\pdv*{\phi}{\xi})\div\vb*{\xi}$. Multiplying both sides of Eq.~\eqref{eq:mu} by $\pdv*{\phi}{\xi}$ and integrating through the interface, we obtain 
\begin{equation}\label{eq:gibbs-thomson}
\tilde{\mu}=\frac{\sigma\kappa}{\Delta\phi}
\end{equation}
\rev{by virtue of Eq.~\eqref{eq:st}. }
Here $\tilde{(\hphantom{\mu})}$ denotes the average in the interface, $\Delta\phi$ the difference of the bulk values of $\phi$ between two sides of the interface, and $\kappa$ the interface curvature, $\kappa=-\div\vb*{\xi}$. 
In simplifying Eq.~\eqref{eq:gibbs-thomson}, we have assumed the homogeneous part of the free energy density $f_0$ is equal on both sides of the interface \cite{Bray1994}, as often modeled by the double-well potential in the form
\begin{equation}
\phi_0(\phi)
\propto
(\phi-\phi_{-\infty})^2(\phi-\phi_{+\infty})^2
\end{equation}
for $\phi=\phi_{\pm\infty}$ as the bulk region away from the interface. Using Eq.~\eqref{eq:gibbs-thomson}, we obtain
\begin{equation}\label{eq:csf}
\tilde{\mu}\grad{\phi}=
\sigma\kappa\frac{\grad\phi}{\Delta\phi},
\end{equation}
which is equivalent to the continuum surface force (CSF) model \cite{Brackbill1992} to approximate the capillary stress term $\sigma\kappa\vb*{\xi}\delta(\xi)$ in volume-of-fluid methods \cite{Tryggvason2011}. 
\rev{Equation \eqref{eq:chella} is therefore an approximate expression to the capillary stress term. In the sharp interface limit, Eq.~\eqref{eq:csf} reduces to Eq.~\eqref{eq:chella}.}

\rev{For the multi-component systems, the free energy model is given by Eq.~\eqref{eq:F_multi}, and the surface tension $\sigma$ is written as
\begin{equation}
\sigma = \sum_{\alpha,\beta} \int\dd{\xi} K_{\alpha\beta} \dv{\phi_{\alpha}}{\xi}\dv{\phi_{\beta}}{\xi}
\end{equation}
instead of Eq.~\eqref{eq:st}. Following the same procedure deriving Eq.~\eqref{eq:gibbs-thomson} for the one-componennt case, we have
\begin{equation}
\sum_{\alpha} \tilde{\mu_{\alpha}}\Delta\phi_{\alpha}
=
\sigma\kappa
- \sum_{\alpha,\beta} \int\dd{\xi}
\frac{K_{\alpha\beta}}{2}\pdv{\phi_{\alpha}}{\xi}\pdv[2]{\phi_{\beta}}{\xi}.
\end{equation}
The capillary stress term for the multi-component systems is now given by
\begin{equation}\label{eq:capillary_multi}
\sigma\kappa\frac{\grad\phi_i}{\Delta\phi_i}
=
\left(
\sum_{\alpha} \tilde{\mu_{\alpha}}\Delta\phi_{\alpha}
+ \sum_{\alpha,\beta} \int\dd{\xi}
\frac{K_{\alpha\beta}}{2}\pdv{\phi_{\alpha}}{\xi}\pdv[2]{\phi_{\beta}}{\xi}
\right)
\frac{\grad\phi_i}{\Delta\phi_i},
\end{equation}
where the choice of the component $i$ is arbitrary since the same interface is shared by all components.
}

\section{Relationship between two formulations under incompressibility condition}\label{sec:confusion}
\rev{%
In numerical simulation of incompressible flows, the pressure is obtained by a projection method to enforce the incompressibility $\div{\vb*{u}}=0$ \cite{kajishima2017}. When Eq.~\eqref{eq:jacqmin} is used instead of Eq.~\eqref{eq:chella} as the capillary stress term in the NS equation for a one-component fluid, by the mathematical identity
\begin{equation}
-\phi\grad{\mu}=\mu\grad{\phi}-\grad(\phi\mu),
\end{equation}
the difference given in the gradient form is absorbed into the pressure calculated by the projection method, and the calculated velocity field turns out to be the same. However, it is emphasized that there is no reason to adopt Eq.~\eqref{eq:jacqmin} to express the capillary stress since $\mu$ should not drive the one-component fluid motion under incompressibility condition. 
}

\rev{%
For multi-component fluids, it is not possible to replace $-\sum_{\alpha}\phi_{\alpha}\grad{\mu_{\alpha}}$ by $\sum_{\alpha}\mu_{\alpha}\grad{\phi_{\alpha}}$ due to the existence of the second term on the RHS of Eq.~\eqref{eq:capillary_multi}.
}

\section{Conclusion}
We have derived two different interfacial stress formulations from the diffuse interface free energy model by two different principles respectively. The two formulations have been both used as the capillary stress formulation in the literature, but we have shown that only one of the two represents the capillary stress. The other formulation represents the fluid stress under the irreversible process, part of which represents the osmotic pressure. 

\appendix
\section{Details on the equations of motion minimizing the Rayleighian}\label{sec:onsager_detail}
\rev{%
Onsager's variational principle is a general framework to describe irreversible processes. It is employed here to derive the equations of fluid motion, where the fluid consists of multiple components. The key quantity is the 
Rayleighian $R$ given by 
\begin{equation}
R=\frac{1}{2}W+\pdv{F}{t}-\int\dd{\vb{r}}p\div{\vb*{u}} \tag{\ref{eq:rayleighian}},
\end{equation}
which is the summation of the dissipation function and the temporal evolution of the system free energy, additionally with the last term to constrain $\div{\vb*{u}}=0$. In the present study, the dissipation function is modeled by the frictions (Eqs.~\ref{eq:w1} and \ref{eq:w2}).
}
The temporal evolution of the system free energy $F$ can be written as 
\begin{align}\label{eq:pfpt}
\pdv{F}{t}
&=\int \dd{\vb*{r}} 
\rev{\sum_{\alpha}} \mu_{\rev{\alpha}}\pdv{\phi_{\rev{\alpha}}}{t}\\
&=-\int \dd{\vb*{r}} 
\rev{\sum_{\alpha}}
\mu_{\rev{\alpha}}[%
\div(\vb*{u}\phi_{\rev{\alpha}})+
\div{\vb*{J}_{\rev{\alpha}}} ],
\end{align}
considering Eq.~\eqref{eq:phi} and denoting $\fdv*{F}{\phi_{\rev{\alpha}}}$ as $\mu_{\rev{\alpha}}$. Since the convective velocity is not considered at this stage, all the fluxes are zero on the system boundaries, and  Eq.~\eqref{eq:pfpt} can be rewritten as
\begin{equation}
\pdv{F}{t}=
\int \dd{\vb*{r}} 
\rev{\sum_{\alpha}}
[(\vb*{u}\phi_{\rev{\alpha}})\vdot\grad{\mu_{\rev{\alpha}}}+\vb*{J}_{\rev{\alpha}}\vdot\grad{\mu_{\rev{\alpha}}}],
\end{equation}
by integrating by parts. Its variation with respect to $\vb*{u}$ is
\begin{equation}
\fdv{(\pdv*{F}{t})}{\vb*{u}}=\rev{\sum_{\alpha}}\phi_{\rev{\alpha}}\grad{\mu_{\rev{\alpha}}}.
\end{equation}
Together with the other terms in $\fdv*{R}{\vb*{u}}$, 
\begin{align}
\fdv{(W_1/2)}{\vb*{u}}&=
-\div{
\left\{
\eta\left[
\grad{\vb*{u}}+(\grad{\vb*{u}})^T
\right]
\right\}
}\\
\fdv{(W_2/2)}{\vb*{u}}&=0\\
-\fdv{}{\vb*{u}}\int\dd{\vb{r}}p\div{\vb*{u}}&=\grad{p},
\end{align}
the vanishing variation of $R$ with respect to $\vb*{u}$ results in the equation of the fluid motion [Eq.~\eqref{eq:stress_onsager}], 
\begin{equation}
\div{
\left\{
\eta\left[
\grad{\vb*{u}}+(\grad{\vb*{u}})^T
\right]
\right\}
}
-\grad{p}
-\rev{\sum_{\alpha}}\phi_{\rev{\alpha}}\grad{\mu_{\rev{\alpha}}}
=0.
\end{equation}
Similarly by taking $\fdv*{R}{\vb*{J}_{\rev{\alpha}}}=0$, we obtain
\begin{equation}
\grad{\mu_{\rev{\alpha}}}=-\frac{\vb*{J}_{\rev{\alpha}}}{M_{\rev{\alpha}}},
\end{equation}
\rev{which is Eq.~\eqref{eq:J}.}

There is a subtlety in evaluating the variation of the free energy evolution in the Rayleghian. When we begin with a different mass conservation equation
\begin{equation}
\pdv{\phi}{t}+\vb*{u}\vdot\grad{\phi_{\rev{\alpha}}}=-\div{\vb*{J}_{\rev{\alpha}}}
\end{equation}
instead of Eq.~\eqref{eq:phi}, which are equivalent under $\div{\vb*{u}}=0$, the \textit{capillary stress} form
\begin{equation}
\fdv{(\pdv*{F}{t})}{\vb*{u}}
=-\rev{\sum_{\alpha}}\mu_{\rev{\alpha}}\grad{\phi_{\rev{\alpha}}}
\end{equation}
is obtained (shown by Qian et al.~\cite{Qian2006a} for one-component fluids). However, as we pointed out in Sec.~\ref{sec:osmotic} it is not reasonable to obtain the capillary stress term as a part of irreversible dynamics, and we conclude that the more general form of the mass conservation equation [Eq.~\eqref{eq:phi}] should be employed. \rev{It is a confusion caused by the ambiguity in the pressure definition when $\div{\vb*{u}}=0$, where it is only the Lagrange multiplier to constrain $\div{\vb*{u}}=0$ as discussed in Sec.~\ref{sec:confusion}.}

\printbibliography

\end{document}